\documentclass[prd,nofootinbib,showpacs,twocolumn]{revtex4-1}
\usepackage{color}
\usepackage{bm}
\usepackage{graphicx}
\usepackage{amsmath}
\usepackage{amssymb}
\usepackage{enumitem}
\usepackage{hyperref}
\usepackage{subfigure}
\usepackage{array}
\usepackage[english]{babel}

\def\be{\begin{equation}}
\def\ee{\end{equation}}
\def\ba{\begin{eqnarray}}
\def\ea{\end{eqnarray}}
\def\l{\left}
\def\r{\right}
\def\f{\frac}

\def\hub{{\mathcal H}}

\def\ie{{\frenchspacing\it i.e. }}

\begin{document}

\title{Effective Field Theory of Cosmic Acceleration:\\
constraining dark energy with CMB data}

\author{ Marco Raveri$^{1,2}$, Bin Hu$^3$, Noemi Frusciante$^{1,2}$, Alessandra Silvestri$^{1,2,4}$}
\affiliation{
\smallskip
$^{1}$ SISSA - International School for Advanced Studies, Via Bonomea 265, 34136, Trieste, Italy \\
\smallskip
$^{2}$ INFN, Sezione di Trieste, Via Valerio 2, I-34127 Trieste, Italy \\
\smallskip
$^{3}$ Institute Lorentz, Leiden University, PO Box 9506, Leiden 2300 RA, The Netherlands \\
\smallskip 
$^{4}$ INAF-Osservatorio Astronomico di Trieste, Via G.B. Tiepolo 11, I-34131 Trieste, Italy}

\begin{abstract}
We introduce EFTCAMB/EFTCosmoMC as publicly available patches to the commonly used CAMB/CosmoMC codes. We briefly describe the structure of the codes, their applicability and main features. To illustrate the use of these patches, we obtain constraints on parametrized \textit{pure} EFT and designer $f(R)$ models, both on $\Lambda$CDM and $w$CDM background expansion histories, using data from {\it Planck} temperature and lensing potential spectra, WMAP low-$\ell$ polarization spectra (WP), and baryon acoustic oscillation (BAO). Upon inspecting  theoretical stability of the models on the given background, we find non-trivial parameter spaces that we translate into {\it viability priors}. We use different combinations of data sets to show their individual effects on cosmological and model parameters. Our data analysis results show that, depending on the adopted data sets, in the $w$CDM background case this {\it viability priors} could dominate the marginalized posterior distributions. 
Interestingly, with {\it Planck}+WP+BAO+lensing data, in $f(R)$ gravity models, we get very strong constraints on the constant dark energy equation of state, $w_0\in(-1,-0.9997)\,\,(95\% {\rm C.L.})$.
\end{abstract}



\maketitle
\section{Introduction}\label{Intro}
Ongoing and upcoming cosmological surveys will map matter and metric perturbations through different epochs with exquisite precision. Combined with geometric probes, they will provide us with a set of independent measurements of cosmological distances and growth of structure. In  the cosmological standard model, $\Lambda$CDM, the rate of linear clustering can be determined from the expansion rate of the Universe; however this consistency relation is generically broken in models of modified gravity (MG) and dark energy (DE), even when they predict the same expansion history as $\Lambda$CDM. Therefore measurements of the growth of structure, such as weak lensing and galaxy clustering, add complementary constraining power to measurements of the expansion history via geometric probes. They can be used to perform consistency tests of $\Lambda$CDM \emph{as well as} to constrain the parameter space of alternative approaches to the phenomenon of cosmic acceleration.

Over the past years there has been a lot of activity in the community to construct  frameworks~\cite{Bertschinger:2006aw,Linder:2007hg,Amendola:2007rr,Zhang:2007nk,Hu:2007pj,Bertschinger:2008zb,Daniel:2008et,Song:2008vm,Skordis:2008vt,Song:2008xd,Zhao:2009fn,Zhao:2010dz,Dossett:2010gq,Song:2010rm,Daniel:2010ky,Pogosian:2010tj,Bean:2010zq,Thomas:2011pj,Baker:2011jy,Thomas:2011sf,Dossett:2011tn,Zhao:2011te,Hojjati:2011xd,Brax:2011aw,Dossett:2012kd,Brax:2012gr,Sawicki:2012re,Baker:2012zs,Amendola:2012ky,Hojjati:2012ci,Silvestri:2013ne,Motta:2013cwa,Asaba:2013xql,Baker:2013hia,Dossett:2013npa,Terukina:2013eqa,Piazza:2013pua,Gergely:2014rna,Hu:2012td,Munshi:2014tua,Hu:2013aqa,Steigerwald:2014ava,Huang:2014fua,Lombriser:2014dua,Tsujikawa:2014mba,Bellini:2014fua} that would allow model-independent tests of gravity with cosmological  surveys like {\it Planck}~\cite{planck}, SDSS~\cite{sdss}, DES~\cite{des}, LSST~\cite{lsst}, Euclid~\cite{euclid}, WiggleZ Dark Energy Survey~\cite{Drinkwater:2009sd,Parkinson:2012vd} and CFHTLenS~\cite{Simpson:2012ra,Heymans:2013fya,Fu:2014loa}. These are generally based on parametrizations of the dynamics of linear 
scalar perturbations, either at the level of the equations of motion, e.g.~\cite{Baker:2013hia}, of solutions of the equations, e.g.~\cite{Bertschinger:2008zb,Pogosian:2010tj,Silvestri:2013ne}, or of the action, e.g.~\cite{Battye:2012eu,Gubitosi:2012hu,Bloomfield:2012ff}, with the general aim of striking a delicate balance among theoretical consistency, versatility and feasibility of the parametrization. We shall focus on the latter approach, in particular on the  effective field theory (EFT) for cosmic acceleration developed in~\cite{Gubitosi:2012hu,Bloomfield:2012ff}  (see~\cite{Weinberg:2008hq,Creminelli:2008wc,Park:2010cw,Cheung:2007st,Jimenez:2011nn,Carrasco:2012cv,Hertzberg:2012qn} for  previous work in the context of inflation, quintessence and large scale structure).  This formalism relies on an action written in unitary gauge and built out of all the operators that are consistent with the unbroken symmetries of the theory, {\it i.e}. time-dependent spatial diffeomorphisms,  and are ordered according  to the power of perturbations and derivatives. At each order in perturbations, there is a finite number of such operators that enter 
the action multiplied 
by time-dependent coefficients commonly referred to as EFT functions. In particular, the background dynamics is determined solely in terms of the three EFT functions multiplying the three background operators, while the general dynamics of linear scalar perturbations is affected by further operators but can still be analyzed in terms of a handful of time-dependent functions. Despite the model-independent construction, there is a precise mapping that can be worked out between the EFT action and the action of any given single scalar field DE/MG model for which there exists a well defined Jordan frame~\cite{Gubitosi:2012hu,Bloomfield:2012ff,Gleyzes:2013ooa,Bloomfield:2013efa}. As such, EFT of cosmic acceleration represents an insightful parametrization of the general quadratic action for single scalar field DE/MG which has the dual quality of offering a unified language to describe a broad range of single field DE/MG models \emph{as well as} a versatile framework for model independent tests of gravity.  For a 
more in depth description of the EFT formalism we refer the reader to the original papers~\cite{Gubitosi:2012hu,Bloomfield:2012ff,Piazza:2013coa}.

In~\cite{Hu:2013twa} we introduced EFTCAMB, a patch which implements the effective field theory formalism of cosmic acceleration into the public Einstein-Boltzmann solver CAMB~\cite{CAMB,Lewis:1999bs}. As such, the code can be used to investigate the implications of the different EFT operators on linear perturbations as well as to study perturbations in any specific dark energy or modified gravity model that can be cast into the EFT language, once the mapping is worked out. Besides its versatility, an important feature of EFTCAMB is that it evolves the full equations without relying on any quasi-static approximation and it still allows for the implementation of specific single field models of DE/MG. Furthermore, as we will briefly review in Section~\ref{Sec:EFTCAMB}, our code has a built-in check of stability of the theory and allows to safely evolve perturbations in models that cross the phantom-divide.
 
We have now completed a modified version of the standard Markov-Chain Monte-Carlo code CosmoMC~\cite{Lewis:2002ah}, that we dubbed  EFTCosmoMC. In combination with the check for stability of the theory embedded in EFTCAMB, it allows to explore the parameter space of models of cosmic acceleration under general viability criteria that are well motivated from the theoretical point of view. It comes with built-in likelihoods for several cosmological data sets.

We illustrate the use of these patches obtaining constraints on different models within the EFT framework using data from
{\it Planck} temperature and lensing potential spectra, WMAP low-$\ell$ polarization (WP) spectra as well as baryon acoustic oscillations (BAO).  In particular we consider designer $f(R)$ models and an EFT linear parametrization involving only background operators, on both a $\Lambda$CDM and $w$CDM bacgkround.

Finally, we are publicly releasing the EFTCAMB and EFTCosmoMC patches at \url{http://wwwhome.lorentz.leidenuniv.nl/~hu/codes/}. We are completing  a set of notes that will guide the reader through the structure of the codes~\cite{HRFS-work-in-progress}.

\section{EFTCAMB}\label{Sec:EFTCAMB}
In this Section we shall briefly review the main features of  EFTCAMB. For a more in depth description of the EFT formalism, the equations evolved by the code and the integration strategy, we refer the reader to~\cite{Hu:2013twa}. 

The EFT formalism for cosmic acceleration is based on the following action written in unitary gauge and Jordan frame
\ba\label{action}
 && S = \int d^4x \sqrt{-g}   \left \{ \frac{m_0^2}{2} \l(1+\Omega\r)R+ \Lambda - a^2c\,\delta g^{00}\r.\nonumber\\
  &&\l.+\frac{M_2^4}{2} (a^2\delta g^{00})^2 - \frac{\bar{M}_1^3 }{2}a^2 \delta g^{00} \delta {K}^\mu_\mu- \frac{\bar{M}_2^2 }{2} (\delta K^\mu_\mu)^2\r. \nonumber\\
  &&\l.- \frac{\bar{M}_3^2 }{2} \delta K^\mu_\nu \delta K^\nu_\mu
+\f{a^2\hat{M}^2}{2}\delta g^{00}\delta R^{(3)}  \right.
\nonumber \\ 
&& \left. + m_2^2 (g^{\mu \nu} + n^\mu n^\nu) \partial_\mu (a^2g^{00}) \partial_\nu(a^2 g^{00})
+ \ldots  \right. \bigg\}\nonumber\\
&& + S_{m} [\chi_i,g_{\mu \nu}],
\ea
where we have used conformal time and $\{\Omega,\Lambda,c,M_2,\bar{M}_1,\bar{M}_2,\bar{M}_3,\hat{M},m_2\}$ are free functions of time which multiply all the operators that are consistent with time-dependent spatial diffeomorphism invariance and are at most quadratic in the perturbations. We will refer to these as EFT functions. The first line of~(\ref{action}) displays only operators that contribute to the evolution of the background. There follow the second order operators that affect only the dynamics of  perturbations, in combination with the background operators. The ellipsis indicate higher order operators which would affect the non-linear dynamics and are not yet included in the code. Finally, $S_m$ is the action for all matter fields, $\chi_i$.

A given single scalar field model of DE/MG, for which there exists a well defined Jordan frame, can be mapped into the formalism~(\ref{action}) as illustrated in~\cite{Gubitosi:2012hu,Bloomfield:2012ff,Gleyzes:2013ooa,Bloomfield:2013efa}. As such, EFT offers a unified language for most of the viable approaches to cosmic acceleration; among other, we shall mention the Horndeski class~\cite{Horndeski:1974wa} which includes quintessence~\cite{Copeland:2006wr}, k-essence~\cite{ArmendarizPicon:1999rj}, $f(R)$, covariant Galileon~\cite{Deffayet:2009wt}, the effective $4D$ limit of DGP~\cite{Nicolis:2008in} to name a few. Furthermore, action~(\ref{action}) represents a parametrized framework to test gravity on large scales via the dynamics of linear cosmological perturbations. To this extent, one can use a designer approach to fix {\it a priori} the background evolution and use the Friedmann equations to determine two of the EFT background functions $\{\Omega,\Lambda,c\}$ in terms of the third one~\cite{Gubitosi:2012hu,Bloomfield:2012ff}. It turns out to be convenient to solve for $c$ and $\Lambda$ in terms of $\Omega$. In the unitary gauge adopted for 
action~(\ref{action}), the extra scalar d.o.f. associated to DE or the modifications of gravity is eaten by the metric. While such set up is convenient to construct the action, in order to analyze the dynamics of perturbations it is convenient to make the scalar d.o.f. explicit. This can be achieved restoring the time-diffeomorphism invariance of the action via the St$\ddot{\text{u}}$ckelberg technique, \ie performing the following infinitesimal time diffeomorphism to the action
\begin{equation}
 \tau \rightarrow \tau + \pi(x^{\mu}),
\end{equation}
and Taylor-expanding the resulting action in $\pi$. The St$\ddot{\text{u}}$ckelberg field, $\pi$, now encodes the departures from the standard cosmological model. At the level of linear scalar perturbations, EFTCAMB evolves the  standard equations for all matter species, a set of modified Einstein-Boltzmann equations and a full Klein-Gordon equation for $\pi$ as described at length in~\cite{Hu:2013twa}. We shall recall that this treatment of the extra dynamics associated to DE/MG, as opposed to an effective fluid approach~\cite{Sawicki:2012re,Bloomfield:2013cyf,Battye:2013ida}, it allows us to maintain a better control of the stability of the theory and to cross the phantom divide.

As described in~\cite{Hu:2013twa}, the multifaceted nature of the EFT formalism is implemented in EFTCAMB, where the background dynamics can be approached with either of the following procedures:
\begin{itemize}[leftmargin=*]
\item \textit{pure} EFT:  in this case one works with a given subset of the operators in~(\ref{action}), possibly all, treating their coefficients as free functions. The background is treated via the EFT designer approach, \ie  a given expansion history is fixed, a viable form for $\Omega$~\cite{Frusciante:2013zop} is chosen and the remaining two background EFT functions are determined via the Friedmann equations. 
The code allows for $\Lambda$CDM and  $w$CDM expansion histories, as well as for the Chevallier-Polarski-Linder (CPL)~\cite{Chevallier:2000qy,Linder:2002et} parametrization of the dark energy equation of state. In addition it offers a selection of functional forms for $\Omega(a)$: the minimal coupling, corresponding to $\Omega=0$; the linear model, that can be thought of as a first order approximation of a Taylor expansion;  power law, inspired by $f(R)$, and exponential ones. There is also the possibility for the user to choose an arbitrary form of $\Omega$ according to any ansatz the user wants to investigate. At the level of perturbations, more operators come into play, each with a free function of time in front of it, and one needs to choose some ans$\ddot{\text{a}}$tze in order to fix their functional form.  To this extent, we adopted the same scheme as for the background function $\Omega$, still providing the possibility to define and use other forms that might be of interest. Of course the possibility 
to set all/some second order EFT functions to zero is included. 
The code evolves the full perturbed equations consistently implemented to account for the inclusion of more than one second order operator per time, ensuring that even more and more complicated models can be studied.
\item \textit{mapping} EFT:  in this case a particular DE/MG model is chosen, the corresponding background equations are solved and then everything is mapped into the EFT formalism~\cite{Gubitosi:2012hu,Bloomfield:2012ff,Gleyzes:2013ooa,Bloomfield:2013efa} to evolve the full EFT perturbed equations. Let us stress that in the \textit{mapping} case, once the background equations are solved, all the EFT functions are completely fixed by the choice of the model.  Built-in to the first code release there are $f(R)$ models for which a designer approach is used for the $\Lambda$CDM, $w$CDM and CPL backgrounds following~\cite{Song:2006ej,Pogosian:2007sw}. In the future other theories will be added to gradually cover the wide range of models included in the EFT framework.
\end{itemize}
In summary, EFTCAMB is a full Einstein-Bolztmann code which exploits the double nature of the EFT framework. As such, it allows to investigate how the different operators entering~(\ref{action}) affect the dynamics of linear perturbations as well as to study a particular DE/MG model, once the mapping to the EFT formalism is determined. Let us stress that EFTCAMB evolves the full dynamical equations on all linear scales both in the \textit{pure} EFT and \textit{mapping} mode,  ensuring that we do not miss out on any potentially interesting dynamics at redshifts and scales that might be within reach of upcoming, wider and deeper, surveys. 

Ultimately, we built a machinery which allows to test gravity and its modifications on large scale in a model independent framework which ideally covers most of the models of cosmological interest by computing cosmological observables which are the two-point auto- and cross-correlations provided by the combination of galaxy clustering, CMB temperature and polarization anisotropy and weak lensing.

\section{EFTCosmoMC: sampling of the parameter space under stability conditions}\label{Sec:EFTCosmoMC}
To fully exploit the power of EFTCAMB we equipped it with a modified version of the standard Markov-Chain Monte-Carlo code CosmoMC~\cite{Lewis:2002ah} that we dubbed EFTCosmoMC. The complete code now allows to explore the parameter space performing comparisons with several cosmological data sets, and it does so with a built-in stability check that we shall discuss in the following.

In the EFT framework, the stability of perturbations in the dark sector can be determined from the equation for the perturbation $\pi$, which is an inhomogeneous Klein-Gordon equation with coefficients that depend both on the background expansion history and the EFT functions~\cite{Gubitosi:2012hu,Bloomfield:2012ff,Hu:2013twa}. Following the arguments of~\cite{Creminelli:2008wc}, in~\cite{Hu:2013twa} we listed general viability requirements in the form of conditions to impose on the coefficients of the equation for $\pi$; these include a speed of sound  $c_s^2\leq 1$, a positive mass $m_\pi^2\geq 0$ and the avoidance of ghost. Furthermore we required a positive non-minimal coupling function, \ie $1+\Omega>0$, to ensure a positive effective Newton constant.

When exploring the parameter space one needs to check the stability of the theory at every sampling point. While this feature at first might seem a drawback, however, it is one of the main advantages of the EFT framework and a virtue of EFTCAMB/EFTCosmoMC. Indeed, as we outlined in~\cite{Hu:2013twa}, checking the stability of the theory ensures not only that the dynamical equations are mathematically consistent and can be reliably numerically solved, but also, perhaps more importantly,  that the underlying physical theory is acceptable. 
This of course is desired when considering specific DE/MG models and, even more, when adopting the \textit{pure} EFT approach. In the latter case indeed, one makes a somewhat arbitrary choice for the functional form of the EFT functions and satisfying the stability conditions will ensure that there is an underlying,  theoretically consistent model of gravity corresponding to that given choice.

Imposing stability conditions generally results in a partition of the parameter space into a stable region and an unstable one. In order not to alter the statistical properties of the MCMC sampler~\cite{Gilks:1999}, like the convergence to the target distribution, when dealing with a partitioned parameter space we implement the stability conditions as priors so that the Monte Carlo step is rejected whenever it would fall in the unstable region. We call these constraints \emph{viability priors} as they represent the degree of belief in a viable underlying single scalar field DE/MG theory encoded in the EFT framework. We would like to stress that they correspond to specific conditions that are theoretically well motivated and, hence, they represent the natural requirements to impose on a model/parametrization. One of the virtues of the EFT framework, and consequently of EFTCAMB/EFTCosmoMC, is to allow their implementation in a straightforward way. 
We shall emphasize that our EFTCosmoMC code 
automatically enforces the {\it viability priors} for every model considered, both in the \textit{pure} and \textit{mapping} EFT approach.
\section{Data sets and Results}
In this Section we shall briefly review the data sets we used and discuss the resulting constraints obtained for some selected  \textit{pure} and \textit{mapping} EFT models on both $\Lambda$CDM and $w$CDM backgrounds. While we will work with models that involve only background operators,  EFTCAMB/EFTCosmoMC are fully equipped to handle also second order operators;  the same procedure that we shall outline here can be followed when the latter are at play.

\subsection{Data sets}
We adopt {\it Planck} temperature-temperature power spectra considering the 9 frequency channels ranging from $30\sim353$ GHz for low-$\ell$ modes ($2\leq\ell<50$)and the $100$, $143$, and $217$ GHz frequency channels  for high-$\ell$ modes ($50\leq\ell\leq2500$) \footnote{\url{http://pla.esac.esa.int/pla/aio/planckProducts.html}}~\cite{Ade:2013kta,Ade:2013zuv}. In addition we include the {\it Planck} collaboration 2013  data release of the full-sky lensing potential map~\cite{Ade:2013tyw}, by using the $100$, $143$, and $217$ GHz frequency bands with an overall significance greater than $25\sigma$. 
The lensing potential distribution is an indicator of the underlying large-scale structure, and as such it is sensitive to the modified growth of perturbations contributing significant constraining power for DE/MG models.

In order to break the well-known degeneracy between the re-ionization optical depth and the amplitude of
CMB temperature anisotropy, we include WMAP low-$\ell$ polarization spectra ($2\leq\ell\leq32$)~\cite{Hinshaw:2012aka}. Finally, we consider the external baryon acoustic oscillations measurements from the 6dFGS ($z= 0.1$)~\cite{Beutler:2011hx}, 
SDSS DR7 (at effective redshift $z_{\rm eff}=0.35$)~\cite{Percival:2009xn,Padmanabhan:2012hf}, 
and BOSS DR9 ($z_{\rm eff}=0.2$ and $z_{\rm eff}=0.35$)~\cite{Anderson:2012sa} surveys to get complementary  constraining power on cosmological distances. 

To explicitly show the effect of individual data sets on the different parameters that we constrain, we adopt three different combinations of data, namely: {\it Planck}+WP; {\it Planck}+WP+BAO; {\it Planck}+WP+BAO+lensing, where with lensing we mean the CMB lensing potential distributions as measured by {\it Planck}. In all cases we assume standard flat priors from CMB on cosmological parameters while we impose the \emph{viability priors} discussed in Section~\ref{Sec:EFTCosmoMC} on model parameters.
\begin{figure*}[!ht]
\includegraphics[scale=0.85]{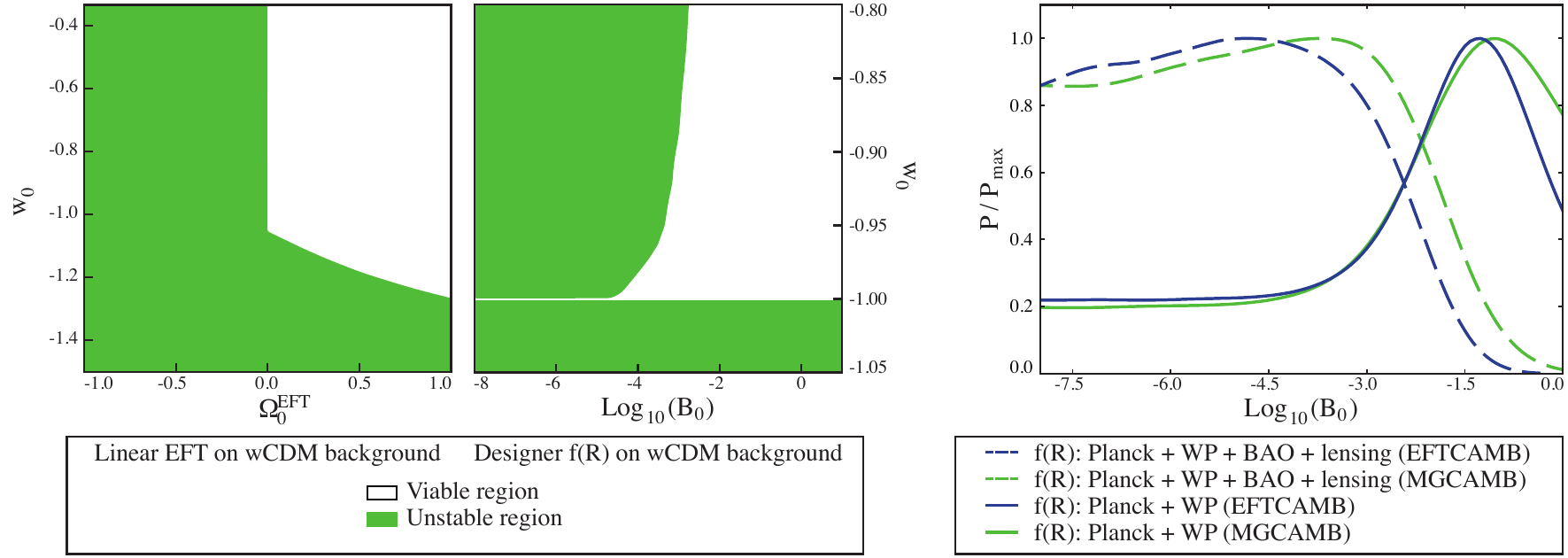}
\caption{\textit{Left panel}: stability regions of linear EFT and designer $f(R)$ models  on a $w$CDM background. The cosmological parameters defining the expansion history are set to their CAMB default values: $H_0=70\,\mbox{Km}/\mbox{s}/\mbox{Mpc}$, $\Omega_b = 0.05$, $\Omega_c=0.22$, $T_{\rm CMB}=2.7255\,\mbox{K}$. \textit{Right panel}: marginalized constraints on ${\rm Log}_{10} (B_0)$ describing designer $f(R)$ models on $\Lambda$CDM background for two data sets differing by CMB lensing and BAO. For each data set we compare the results obtained with EFTCAMB with those obtained by MGCAMB~\cite{Zhao:2008bn,Hojjati:2011ix} for the same designer $f(R)$ models.}
\label{fig:Stab_MGvsEFT}
\end{figure*}
\subsection{Linear EFT model\label{Sec:linear_model}} 
We start our exploration of CMB constraints on DE/MG theories with a \textit{pure} EFT model. We adopt the designer approach choosing two different models for the expansion history, the $\Lambda$CDM one and the $w$CDM one (corresponding to a constant dark energy equation of state). As we reviewed in Section~\ref{Sec:EFTCAMB}, after fixing the background expansion history one can use the Friedmann equations to solve for two of the three EFT background functions in terms of the third one; as it is common, we use this to eliminate $\Lambda$ and $c$. We are then left with $\Omega$ as a free background function that will leave an imprint only on the behaviour of perturbations. We assume the following functional form:
\begin{align}
\Omega(a) = \Omega_0^{\rm EFT} \, a \,,
\end{align}
which can be thought of as a first order approximation of a Taylor expansion in the scale factor. We set to zero the coefficients of all the second order EFT operators. In the remaining we refer to this model as the linear EFT model.

Before proceeding with parameter estimation, it is instructive to study the shape of the viable region in the parameter space of the model. As we discussed in Section~\ref{Sec:EFTCosmoMC}, the check on the stability of any given model is a built-in feature of EFTCAMB/EFTCosmoMC, so that the user does not need to separately perform such an investigation prior to implementing the model in the code. Nevertheless, in some cases it might be useful to look at the outcome of such analysis as one can learn interesting things about the model/parametrization under consideration. Let us briefly discuss the stability of the linear EFT model.

In the case of a $\Lambda$CDM expansion history, it is easy to show that all the stability requirements that we listed in~\cite{Hu:2013twa}, and reviewed in Section~\ref{Sec:EFTCosmoMC}, imply the following \textit{viability prior}:
\begin{align}
\Omega_0^{\rm EFT} \geq 0 \,.
\end{align}

On the other hand, the case of a $w$CDM expansion history can not be treated analytically 
so we used our EFTCAMB code along with a simple sampling algorithm, 
included in the code release, to explore the stability of the model in the parameter space. 
We varied the parameters describing the dark sector physics while keeping fixed all the other cosmological parameters. 
The result is shown in Figure~\ref{fig:Stab_MGvsEFT} and includes interesting information on the behaviour of this model. 
First of all, also in this case the stable region correspond to $\Omega_0^{\rm EFT}>0$; 
furthermore it is possible to have a viable gravity model with $w_0 < -1$, 
although in this case $\Omega_0^{\rm EFT}$ needs to acquire a bigger and bigger value to stabilize perturbations in the dark sector. 
Finally, we see that if $\Omega_0^{\rm EFT}=0$ we recover the result, found in the context of quintessence models~\cite{Zlatev:1998tr}, that $w_0 >-1$. This case corresponds, in fact, to minimally coupled quintessence models with a  potential that is crafted so that the resulting expansion history mimics that of a $w$CDM model. 

For the $\Lambda$CDM background case, the 1D marginalized posterior distributions, 
obtained with the three different data compilations discussed above, are shown in Figure~\ref{fig:Marginal1D} (a).
The corresponding marginalized statistics are summarized in Table~\ref{Tab:stat} (a). 
We find that the three different data compilations produce similar results, with 
{\it Planck}+WP+BAO+lensing giving:
\be
\label{LiEFT_LCDM:Omega0}
\Omega_0^{\rm EFT}<0.061\;\;\;(95\%{\rm C.L.})\; .
\ee

\begin{figure*}[ht]
\includegraphics[scale=0.918]{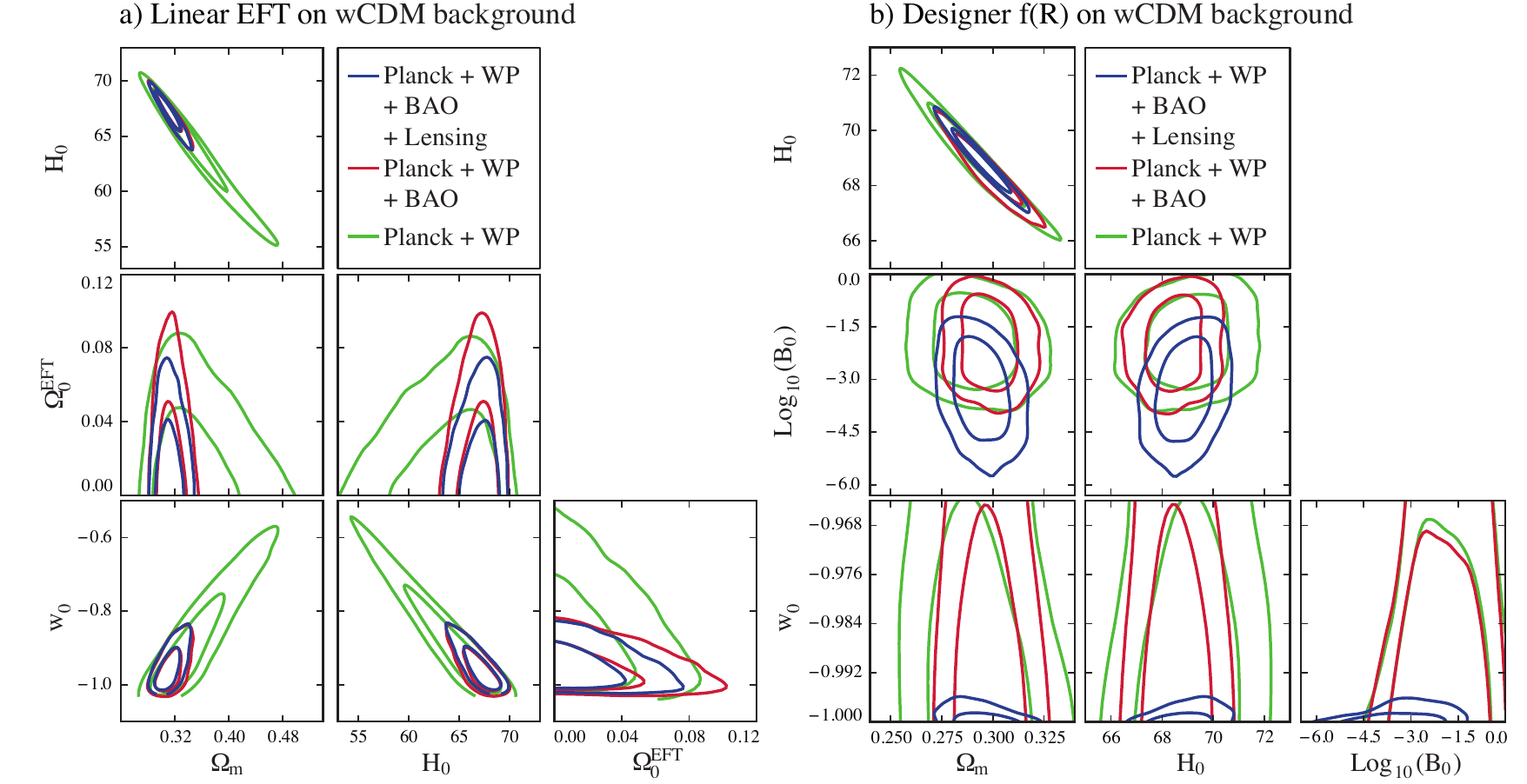}
\caption{ $68\%$ and $95\%$ confidence regions on combinations of cosmological parameters for linear \textit{pure} EFT and designer $f(R)$ models on $w$CDM background. Different combinations of observables are indicated with different colors.}
\label{fig:Marginal2D}
\end{figure*}

Next we consider a $w$CDM expansion history, characterized by an equation of state for dark energy constant in time, $w_0$, but different from $-1$. 
Upon inspecting Figure~\ref{fig:Marginal1D} (b) one can notice that the marginalized posterior distributions of 
($\Omega_m,\Omega_{\Lambda},H_0,w_0$) obtained from {\it Planck}+WP data are significantly skewed, 
\ie their right tail goes to zero much more sharply than the left one. The situation changes significantly when one adds BAO data.
This is due to the combination of two effects. On one hand, when BAO data are not included, the constraints on ($\Omega_m,\Omega_{\Lambda},H_0,w_0$) are relatively loose since one is lacking the complementary high precision information on the expansion history. In other  words, the gain/loss of likelihood value in the vicinity of best-fit points is not very significant, so the sampling points of cosmological parameters broadly spread around their central values.  In this case, the stability requirements on $\Omega_0^{\rm EFT}$ and $w_0$ dominate over the data constraining power. On the other hand, as shown in the left panel of Figure~\ref{fig:Stab_MGvsEFT},  the viable region in the space ($\Omega_0^{\rm EFT},w_0$) for 
the linear EFT model on a $w$CDM background covers mostly $w_0>-1$, \ie it is highly asymmetric in the range around $w_0=-1$. This explains the asymmetry in the posterior distribution of $w_0$ since the marginalized posterior distribution in Monte-Carlo integration algorithms follows the number of projected sampling points in the given volume. Furthermore, from the left panel of Figure~\ref{fig:Marginal2D} (green curve), one can see that ($\Omega_m,\Omega_{\Lambda},H_0$) are degenerate, as expected, with $w_0$ and this explains while their posterior distributions are skewed as well. As soon as complementary measurements of cosmological distances, such as BAO,  are added to the data sets, the constraining power is strong enough and the posterior distributions become more symmetric; indeed BAO data significantly helps to localize the confidence regions close to $w_0\sim-1$, making the posterior distribution less affected by the global profile of {\it viability priors}.

Finally, from the left panel of Figure~\ref{fig:Marginal2D} we can see that the degeneracy of $\Omega_0^{\rm EFT}$ with the other parameters is not very significant after adding BAO data (blue and red curves). As a result the bounds on $\Omega_0^{\rm EFT}$ remain at the same level of those obtained for a $\Lambda$CDM background. With {\it Planck}+WP+BAO+lensing data we obtain:
\be
\label{LiEFT_wCDM:Omega0}
\Omega_0^{\rm EFT}<0.058\;\;\;(95\%{\rm C.L.})\;.
\ee
One can notice that the addition of lensing data does not significantly improve the constraint on $\Omega_0^{\rm EFT}$ in neither the $\Lambda$CDM nor the $w$CDM case.

\subsection{$f(R)$ gravity\label{Sec:f_R}}
The simplicity of its theoretical structure and the representative phenomena  to which it leads at the level of growth of structure, have long made $f(R)$ gravity a popular model of modified gravity. We refer 
the reader to~\cite{Song:2006ej,Bean:2006up,Pogosian:2007sw,DeFelice:2010aj} for detailed discussions of the cosmology of $f(R)$ models. Here we shall briefly review the main features that are of interest for our analysis. 

We consider the following action in Jordan frame
\be\label{action_fR}
S=\int d^4x \sqrt{-g} \l[R+f(R)\r]+S_m \,,
\ee
where $f(R)$ is a generic function of the Ricci scalar and the matter sector is minimally coupled to gravity. 
The higher order nature of the theory translates into having an extra scalar d.o.f. which can be identified with $f_R\equiv df/dR$, commonly dubbed the \emph{scalaron}~\cite{Starobinsky:2007hu}. As explained in~\cite{Gubitosi:2012hu}, these models can be mapped into the EFT formalism via the following matching:
\begin{align}\label{fR_matching}
\Lambda=\frac{m_0^2}{2}\l[f- Rf_R \r] \,\,;\hspace{0.5cm} c=0 \,\,;\hspace{0.5cm} \Omega=f_R \,.
\end{align}
Within the EFT language, the role of the perturbation to the scalaron, $\delta f_R$, is played by the St$\ddot{\text{u}}$ckelberg field which, in the $f(R)$ case, corresponds to $\pi=\delta R/R$~\cite{Gubitosi:2012hu}.

It is well known that given the higher order of the theory, 
it is possible to reproduce any given expansion history by an appropriate choice of the $f(R)$ function
~\cite{Song:2006ej,Pogosian:2007sw}. 
In other words, $f(R)$ models can be treated with the so called \emph{designer} 
approach which consists in fixing the expansion history and then using the 
Friedmann equation as a second order differential equation for $f[R(a)]$. Generically one finds a family of viable models that reproduce the given expansion; the latter are commonly labelled by the boundary condition at present time, $f_R^0$. 
Equivalently, they can be parametrized by the present value, $B_0$, of the Compton wavelength of the scalaron in Hubble units
\begin{equation}
\label{ComptonWave}
B=\frac{f_{RR}}{1+f_R}\frac{\hub \dot{R}}{\dot{\hub}-\hub^2}\,,
\end{equation} 
where dots indicate derivatives w.r.t. conformal time. $B_0$ and can be related to $f_R^0$ and one approximately has $B_0\approx-6f_R^0$ in $\Lambda$CDM case. Let us recall that the heavier the scalaron the smaller $B_0$ and $|f_R^0|$. 

Finally, as discussed in~\cite{Amendola:2006we,Song:2006ej,Pogosian:2007sw, Hu:2007nk}, $f(R)$ models need to satisfy certain conditions of stability and consistency 
with local tests of gravity~\cite{Pogosian:2007sw}; the latter can be inferred from the conditions discussed in Section~\ref{Sec:EFTCosmoMC}
once the matching~(\ref{fR_matching}) is implemented. 

As described at length in our previous work~\cite{Hu:2013twa}, EFTCAMB treats the background of $f(R)$ gravity with a built-in designer routine that is specific to these models and can handle $\Lambda$CDM, $w$CDM and CPL backgrounds. Furthermore, the viability of the reconstructed model is automatically checked by the code via the procedure described in Section~\ref{Sec:EFTCosmoMC}.

Like in the \textit{pure} EFT case, it proves very instructive to investigate the shape of the parameter space as dictated by the stability conditions of Section~\ref{Sec:EFTCosmoMC}. For the designer $f(R)$ model on $\Lambda$CDM background it is easy to show that the latter reproduce the known result that in order to have a positive mass of the scalaron $B_0$ should be greater than zero. 
It is much more interesting to investigate the shape of the parameter space for $f(R)$ models mimicking a $w$CDM background expansion history. We do it numerically, through the built-in routine of EFTCAMB, and we show the results in Figure~\ref{fig:Stab_MGvsEFT}.
The first noticeable feature is that for $w$CDM models the value of the equation of state of dark energy can not go below $-1$, which is consistent with what was found in~\cite{Pogosian:2007sw}.
The second one is that the parameter $B_0$ controls the limit to GR of the theory \ie when $B_0$ gets smaller the expansion history is forced to go back to that of the $\Lambda$CDM model in order to preserve a positive mass of the scalaron.
The converse is not true and this same feature do not appear in \textit{pure} EFT models where the $\Omega_0^{\rm EFT}=0$ branch contains viable theories and corresponds to the wide class of minimally coupled quintessence models.

In what follows, we shall first investigate the constraints on $B_0$ in models reproducing $\Lambda$CDM background, performing also a comparison with analogous results obtained using MGCAMB~\cite{Zhao:2008bn,Hojjati:2011ix}. We will then move to study constraints on designer models on a $w$CDM background, which is a novel aspect of our work. 

In the right panel of Figure~\ref{fig:Stab_MGvsEFT}, we compare the 1D marginalized posterior distributions of ${\rm Log}_{10} B_0$ from our EFTCAMB to those from MGCAMB. Overall there is good agreement between the two results. 
Moreover, one can notice that generally the constraints obtained with EFTCAMB are a little bit tighter than those obtained with MGCAMB. This is because in the latter code $f(R)$ models are treated with the quasi-static approximation which looses out on some of the dynamics of the scalaron~\cite{Hojjati:2012rf}, which is instead fully captured by our full Einstein-Boltzmann solver, as discussed already in~\cite{Hu:2013twa}.

The detailed 1D posterior distributions and corresponding marginalized statistics are summarized in Figure~\ref{fig:Marginal1D} (c) and Table~\ref{Tab:stat} (c) and they are consistent with previous studies employing the quasi-static approximations~\cite{Dossett:2014oia}.
The right panel of Figure~\ref{fig:Stab_MGvsEFT} and Figure~\ref{fig:Marginal1D} (c) show that lensing data add a significant consrtaining power on $B_0$. This is because {\it Planck} lensing data are helpful in breaking the degeneracy between $\Omega_m$ and $B_0$ which affect the lensing spectrum in different ways. Indeed, in $f(R)$ gravity the growth rate of linear structure is enhanced by the modifications, hence the amplitude of the lensing potential spectrum is amplified whenever $B_0$ is different than zero (see our previous work~\cite{Hu:2013twa}); however,  the background angular diameter distance is not affected by $B_0$, so the position of the lensing potential spectrum is not shifted horizontally. On the other hand, $\Omega_m$ affects both the background and linear perturbation so that both the amplitude and position of the peaks of the lensing potential are sensitive to it.

Similarly to what happens in the linear EFT model, $f(R)$ gravity shows some novel features in the case of a $w$CDM background. Once again we find a non-trivial likelihood profile of ${\rm Log}_{10}B_0$ (see Figure~\ref{fig:Marginal1D} (d)) for all the three data compilations, with the shape of the marginalized posterior distribution of ${\rm Log}_{10}B_0$ being dominated by the shape of the stable region. In the middle panel of Figure~\ref{fig:Stab_MGvsEFT} one can see indeed that when $B_0$ tends to smaller values, \ie the theory tends to GR, the stable regions becomes narrower and narrower, with  a tiny tip pointing to the GR limit. 
Since the width of this tip is so narrow compared with the current capability of parameter estimation from {\it Planck} data, 
the gains of likelihood of the sampling points inside this parameter throat are not significant, \ie they are uniformly sampled in the throat.
Hence, even though the full data set has a very good sensitivity to $B_0$, the marginalized distribution of ${\rm Log}_{10}B_0$ is dominated by the volume of the stable region in the parameter space. A complementary consequence of the shape of the stable region in the $(B_0,w_0)$ space is the fact that when $B_0$ tends to zero, $w_0$ is driven to $-1$. In other words the stability conditions induce a strong correlation between $B_0$ and $w_0$ and, as a consequence, in $f(R)$ models, no matter in the $\Lambda$CDM or $w$CDM background case, the GR limit is effectively controlled by a single parameter, \ie $B_0$.

Finally one can notice that the bound on $w_0$ with {\it Planck} 
lensing data is quite stringent compared to those without lensing, namely:
\begin{eqnarray}
\label{w0:lensing}
w_0&\in&(-1,-0.94)\; (95\%{\rm C.L.})\;{\rm without~lensing},\nonumber\\
\label{w0:nolensing}
w_0&\in&(-1,-0.9997)\; (95\%{\rm C.L.})\;{\rm with~lensing}.
\end{eqnarray}
We argue that this stringent constraint actually is a consequence of the combination of  the strong correlation between $B_0$ and $w_0$ induced by the viability prior, as discussed above, and the sensitivity of lensing data to $B_0$, that we capture well with our code.  As shown in~\cite{Hu:2013aqa}  {\it Planck} lensing data is very sensitive to MG parameters such as $B_0$; indeed, in our analysis with {\it Planck}+WP+BAO+lensing data we get
\be
\label{B0:wCDM}
{\rm Log}_{10}B_0=-3.35^{+1.79}_{-1.77}\;\;\;(95\%{\rm C.L.})\;.
\ee
Furthermore, from Figure~\ref{fig:Marginal2D} (b), one can see that the ellipse in the  $({\rm Log}_{10}B_0,w_0)$ space  corresponding to {\it Planck}+WP+BAO+lensing data (blue) is orthogonal to those without lensing (red and green). In other words, when lensing data is included, ${\rm Log}_{10}B_0$ and $w_0$ display a degeneracy which propagates the stringent constraint on
the scalar Compton wavelength from CMB lensing data to $w_0$. 
Besides this, we do not find other remarkable degeneracies between $B_0$ and standard cosmological parameters. 
%
\section{Conclusions}
Solving the puzzle of cosmic acceleration is one of the major challenges of modern cosmology. In this respect, cosmological surveys will provide a large amount of high quality data allowing to test gravity on large scales with unprecedented accuracy.
The effective field theory framework for cosmic acceleration will prove useful in performing model independent tests of gravity as well as in testing specific theories as, while being a parametrization of the quadratic action,  it preserves a direct link to a wide range of DE/MG models. Indeed, any single scalar field DE/MG models with a well defined Jordan frame can be cast into the EFT language and most of the cosmological models of interest fall into this class. 

In a previous work~\cite{Hu:2013twa}, we presented the implementation of the EFT formalism into the Einstein-Boltzmann solver CAMB~\cite{CAMB}, resulting into  EFTCAMB. The latter code has several virtues: it allows for the implementation of the parameterized EFT framework in what we dub the \textit{pure} EFT approach, and for the implementation of specific DE/MG models in the \textit{mapping} mode;  it does not rely on any quasi-static approximation, but rather it  implements the full  perturbative equations on all linear scale for both the \textit{pure} and the \textit{mapping} case ensuring that no potentially interesting physics is lost; it has a built-in check of the stability conditions of perturbations  in the dark sector in order to guarantee that the underlying gravitational theory is viable; it enables to choose among different expansion histories, namely  $\Lambda$CDM, $w$CDM and CPL backgrounds, naturally allowing phantom-divide crossings.

In the present work, we equipped EFTCAMB with a modified version of CosmoMC, that we dubbed EFTCosmoMC, creating a bridge between the EFT parametrization of the dynamics of perturbations and observations. EFTCosmoMC allows to practically perform tests of gravity and get constraints analyzing the cosmological parameter space with, in its current version, data sets, such as \textit{Planck}, WP, BAO and \textit{Planck} lensing. Further data sets, mainly from large-scale structure surveys, will be included in the near future. 

As discussed in Section~\ref{Sec:EFTCosmoMC}, exploring the parameter space requires a step by step check of the stability of the theory.  We implemented the resulting stability conditions as  {\it viability priors} that makes the Monte Carlo step be rejected whenever it would fall in the unstable region of the parameter space. The latter procedure, in our view, represents a clean and natural way to impose priors on parameters describing the dark sector.

To illustrate the use of the EFTCAMB/EFTCosmoMC package, we have derived constraints on two different classes of models, namely a \textit{pure} linear EFT model and a  \textit{mapping} designer $f(R)$. We used three different combinations of \textit{Planck}, WP, BAO and CMB lensing data sets to  show their different effects on constraining the parameter space. For both models we have adopted the designer approach built-in in EFTCAMB and have considered the case of a $\Lambda$CDM as well as of a $w$CDM background.

For the linear EFT model, we have derived bounds on the only model parameter, i.e  the present value of the conformal coupling functions $\Omega_0^{\rm EFT}$, as described in Section~\ref{Sec:linear_model}.
In the case of a $\Lambda$CDM background, we have found that the latter needs to satisfy $\Omega_0^{\rm EFT}\geq 0$ as a viability condition and with {\it Planck}+WP+BAO+lensing data we get a bound of $\Omega_0^{\rm EFT} < 0.061$ (95\%C.L.) (the three different data compilations give similar results).  For the $w$CDM expansion history, the outcome of the stability analysis is shown in Figure~\ref{fig:Stab_MGvsEFT}; specifically,  there is a stable region in parameter space where the dark energy equation of state can be smaller than $-1$ as long as the corresponding value of $\Omega_0^{\rm EFT}$ is high enough to stabilize perturbations in the dark sector; finally, the value $\Omega_0^{\rm EFT}=0$  corresponds to a minimally coupled model and requires $w_0 >-1$, like in the case of quintessence.  The combined  bound on $\Omega_0^{\rm EFT}$ with {\it Planck}+WP+BAO+lensing data gives $\Omega_0^{\rm EFT} < 0.058$ (95\%C.L.). 

\begin{figure*}[htb!]
\centering
\includegraphics{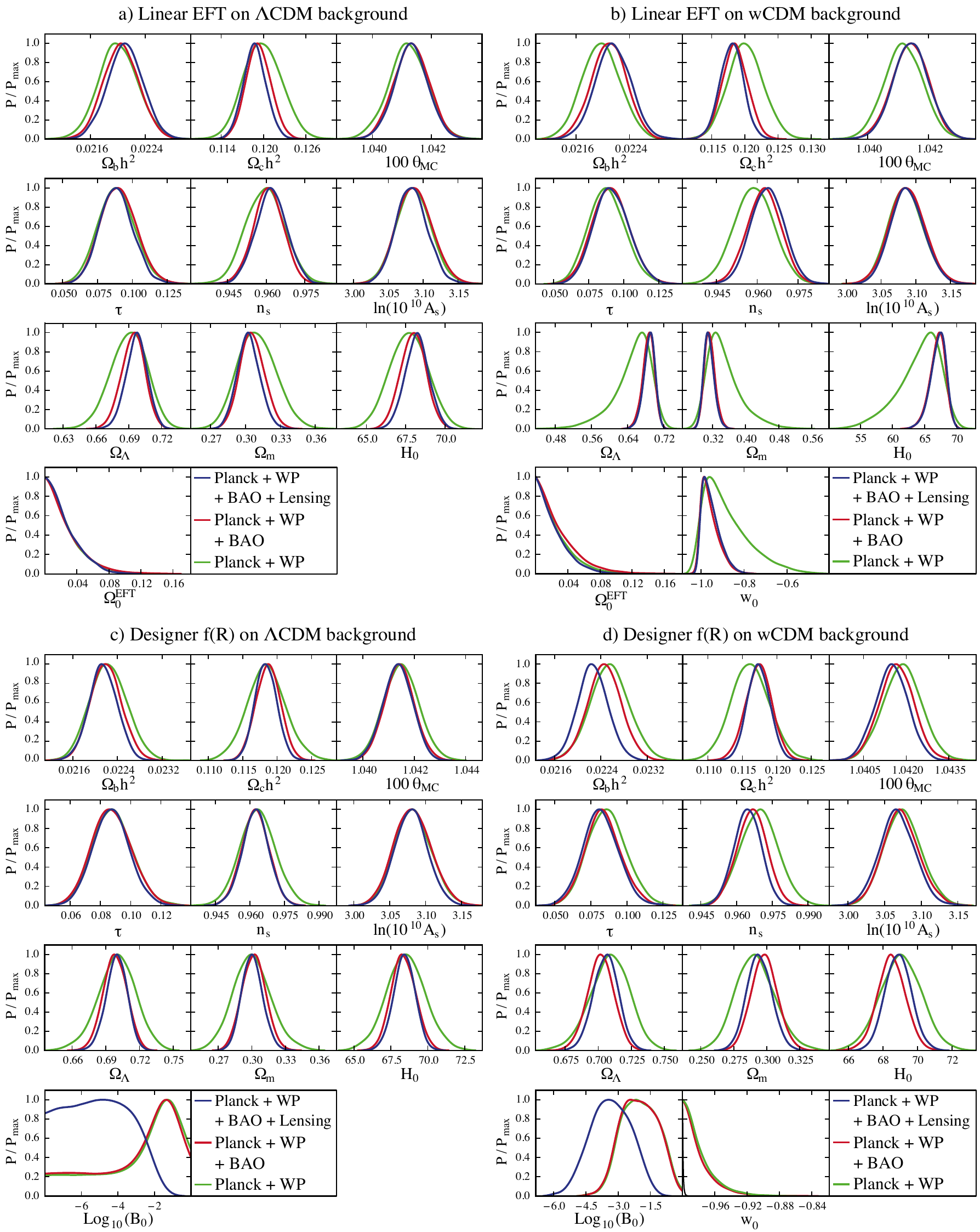}
\caption{1D Marginalized posterior distributions of cosmological and model parameters for \textit{pure} linear EFT (top) and $f(R)$ (bottom) models both on a $\Lambda$CDM (left) and $w$CDM (right) background. Different colors represent different combinations of cosmological data sets.}
\label{fig:Marginal1D}
\end{figure*}

Finally, we have investigated designer $f(R)$ models on $\Lambda$CDM/$w$CDM backgrounds, in terms of constraints on the model parameter $B_0$, as described in Section~\ref{Sec:f_R}. For the $\Lambda$CDM case we also compared our results to those that we obtained with the quasi-static treatment of these models via MGCAMB~\cite{Zhao:2008bn,Hojjati:2011ix}. 
The two treatments give results that are in good agreement, with bounds from EFTCAMB/EFTCosmoMC  being a little tighter thanks to the full treatment of the dynamics of perturbations.
On $w$CDM background  we have found a non-trivial likelihood profile of $\text{Log}_{10}B_0$ (see Figure~\ref{fig:Marginal1D} (d)) for all the three data compilations and the shape of the marginalized posterior distribution in this case strongly reflects that of the viable region in parameter space. In the $w$CDM background, with {\it Planck}+WP+BAO+lensing data we get ${\rm Log}_{10}B_0=-3.35^{+1.79}_{-1.77}$ (95\% C.L.).
The bounds on $w_0$ with \textit{Planck} lensing data ($w_0\in(-1,-0.9997)$ (95\% C.L.)) are quite stringent compared to those without this data set ($w_0\in(-1,-0.94)$ (95\% C.L.)) due to the high constraining power of lensing measurements on $B_0$ and the strong correlation between $w_0$ and $B_0$ via the \textit{viability prior}.

Within the \textit{pure} EFT approach, the code we are releasing contains already all operators relevant for the dynamics of linear perturbations. In particular, while in this work we showed results for a model involving only background operators, we shall stress that  the code is fully functional for second order operators too. 
In the future we will add some third order operators to study mildly non-linear scales. 
As for the \textit{mapping} mode, the code currently allows for a \emph{full} treatment of $f(R)$ models via a specific designer approach that can handle both $\Lambda$CDM, $w$CDM and CPL backgrounds. 
We will gradually implement the \textit{mapping} procedure for many other single scalar field DE/MG models which are of relevance for cosmological tests.
Finally, more data sets will be added into EFTCosmoMC allowing to test gravity with the most recent data releases with a particular emphasis toward large scale structure observations in view of surveys such as Euclid~\cite{euclid}. 

The complete EFTCAMB/EFTCosmoMC bundle is now publicly available at \url{http://wwwhome.lorentz.leidenuniv.nl/~hu/codes/}.
\acknowledgments
We are grateful to Carlo Baccigalupi, Matteo Calabrese, Luca Heltai, Matteo Martinelli and Riccardo Valdarnini for useful conversations. 
AS is grateful to Alireza Hojjati, Levon Pogosian and Gong-Bo Zhao for their previous and ongoing collaboration on related topics.
BH is supported by the Dutch Foundation for Fundamental Research on Matter (FOM).
NF acknowledges partial financial support from the European Research Council under the European Union's Seventh 
Framework Programme (FP7/2007-2013) / ERC Grant Agreement n.~306425 ``Challenging General Relativity''. 
AS acknowledges support from a SISSA Excellence Grant. MR and AS acknowledge partial support from the INFN-INDARK initiative.


\setlength\tabcolsep{1pt}
\begin{table*}[htb!]
\footnotesize
\centering
\subtable[\,Mean values and $68\%$ (or $95\%$) confidence limits for primary/derived parameters for a linear EFT model on $\Lambda$CDM/$w$CDM background.]{%
\begin{tabular}{|l||c|c|c||c|c|c|}
\hline
\multicolumn{1}{|l||}{  }&
\multicolumn{3}{c||}{Linear EFT+$\Lambda$CDM} &
\multicolumn{3}{c|}{Linear EFT+$w$CDM}\\
\hline
\hline
                               &  {\it Planck}+WP          & {\it Planck}+WP+BAO     &\parbox[t]{2.5cm}{{\it Planck}+WP+\\BAO+lensing}           &  {\it Planck}+WP        & {\it Planck}+WP+BAO              & \parbox[t]{2.5cm}{{\it Planck}+WP+\\BAO+lensing} \\ 
\hline
Parameters                     & mean $\pm$ $68\%$ C.L.    & mean $\pm$ $68\%$ C.L.  & mean $\pm$ $68\%$ C.L.  & mean $\pm$ $68\%$ C.L. & mean $\pm$ $68\%$ C.L.  & mean $\pm$ $68\%$ C.L. \\ \hline
$100\Omega_b h^2$              & 2.201$\pm$0.028           & 2.205$\pm$0.025         & 2.211$\pm$0.025         & 2.198$\pm$0.028        & 2.209$\pm$0.026          & 2.216$\pm$0.026 \\
$\Omega_c h^2$                 & 0.1199$\pm$0.0026         & 0.1193$\pm$0.0017       & 0.1188$\pm$0.0016       & 0.1201$\pm$0.0026      & 0.1185$\pm$0.0019        & 0.1180$\pm$0.0018 \\
$100\theta_{\rm MC}$           & 1.04121$\pm$0.00063       & 1.04133$\pm$0.00058     & 1.04132$\pm$0.00055     & 1.04119$\pm$0.00062    & 1.04141$\pm$0.00058     & 1.04142$\pm$0.00058 \\
$\tau$                         & 0.089$\pm$0.013           & 0.090$\pm$0.013         & 0.088$\pm$0.012         & 0.088$\pm$0.013        & 0.091$\pm$0.013          & 0.091$\pm$0.012 \\
$n_s$                          & 0.9596$\pm$0.0073         & 0.9608$\pm$0.0057       & 0.9619$\pm$0.0059       & 0.9588$\pm$0.0071      & 0.9625$\pm$0.0060        & 0.9637$\pm$0.0060 \\
${\rm Log}(10^{10} A_s)$            & 3.086$\pm$0.024           & 3.088$\pm$0.025         & 3.084$\pm$0.022         & 3.086$\pm$0.025        & 3.088$\pm$0.025          & 3.088$\pm$0.023 \\
$\Omega_0^{\rm EFT}$           & $<0.066$ ($95\%$C.L.)       & $<0.072$ ($95\%$C.L.)     & $<0.061$ ($95\%$C.L.)     & $<0.065$ ($95\%$C.L.)    & $<0.076$ ($95\%$C.L.)     & $<0.058$ ($95\%$C.L.) \\
$w_0$                          & $-$                       & $-$                     & $-$                     & $-0.88^{+0.21}_{-0.14}$ ($95\%$C.L.) & $-0.96^{+0.09}_{-0.06}$ ($95\%$C.L.) & $-0.95^{+0.08}_{-0.07}$ ($95\%$C.L.) \\
\hline
$\Omega_{m}$                   & 0.310$\pm$0.016           & 0.306$\pm$0.010         & 0.3028$\pm$0.0096       & 0.349$\pm$0.041        & 0.314$\pm$0.013          & 0.312$\pm$0.013 \\
$H_0$                          & 67.71$\pm$1.20            & 67.99$\pm$0.79          & 68.22$\pm$0.75          & 64.10$\pm$3.26         & 66.99$\pm$1.22           & 67.08$\pm$1.21 \\
\hline
\hline
$\chi^2_{\rm min}/2$           & 4902.799                  & 4904.074                & 4908.849                & 4902.921               & 4903.957                 & 4908.846 \\
\hline
\end{tabular}
}\vspace{1cm}
\subtable[\,Mean values and $68\%$ (or $95\%$) confidence limits for primary/derived parameters for a designer $f(R)$ model on $\Lambda$CDM/$w$CDM background.]{%
\begin{tabular}{|l||c|c|c||c|c|c|}
\hline
\multicolumn{1}{|l||}{}&
\multicolumn{3}{c||}{$f(R)$+$\Lambda$CDM} &
\multicolumn{3}{c|}{$f(R)$+$w$CDM} \\
\hline
\hline
                           &  {\it Planck}+WP      & {\it Planck}+WP+BAO & \parbox[t]{2.5cm}{{\it Planck}+WP+\\BAO+lensing}   &  {\it Planck}+WP    & {\it Planck}+WP+BAO   & \parbox[t]{2.5cm}{{\it Planck}+WP+\\BAO+lensing} \\ 
\hline
Parameters                     & mean $\pm$ $68\%$ C.L.   & mean $\pm$ $68\%$ C.L.  & mean $\pm$ $68\%$ C.L.   &  mean $\pm$ $68\%$ C.L.    & mean $\pm$ $68\%$ C.L.  & mean $\pm$ $68\%$ C.L. \\ \hline
$100\Omega_b h^2$              & 2.224$\pm$0.033          & 2.220$\pm$0.027         & 2.214$\pm$0.025          & 2.255$\pm$0.033           & 2.246$\pm$0.029         & 2.226$\pm$0.026 \\
$\Omega_c h^2$                 & 0.1185$\pm$0.0027        & 0.1187$\pm$0.0017       & 0.1184$\pm$0.0016        & 0.1162$\pm$0.0027         & 0.1174$\pm$0.0019      & 0.1174$\pm$0.0016 \\
$100\theta_{\rm MC}$           & 1.04149$\pm$0.00067      & 1.04142$\pm$0.00057     & 1.04136$\pm$0.00056      & 1.04186$\pm$0.00066       & 1.04166$\pm$0.00060  & 1.04149$\pm$0.00056 \\
$\tau$                         & 0.088$\pm$0.013          & 0.087$\pm$0.013         & 0.087$\pm$0.012          & 0.086$\pm$0.013           & 0.084$\pm$0.012         & 0.082$\pm$0.012 \\
$n_s$                          & 0.9634$\pm$0.0076        & 0.9624$\pm$0.0058       & 0.9625$\pm$0.0057        & 0.9695$\pm$0.0078         & 0.9665$\pm$0.0062      & 0.9647$\pm$0.0057 \\
${\rm Log}(10^{10} A_s)$            & 3.083$\pm$0.025          & 3.082$\pm$0.025         & 3.080$\pm$0.022          & 3.075$\pm$0.025           & 3.072$\pm$0.024         & 3.067$\pm$0.024 \\
${\rm Log}_{10}B_0$                 & $<0.0$ (\footnote{No significant upper bound found in the parameter range we investigated.})        & $<0.0$ ($^a$)     & $<-2.37$ ($95\%$C.L.)      & $-1.97^{+1.61}_{-1.52}$ ($95\%$C.L.)    & $-2.01^{+1.60}_{-1.51}$ ($95\%$C.L.)       & $-3.35^{+1.79}_{-1.77}$ ($95\%$C.L.)\\
$w_0$                          & $-$                      & $-$                     & $-$              & $(-1,-0.94)$ ($95\%$C.L.)  & $(-1,-0.94)$ ($95\%$C.L.) & $(-1,-0.9997)$ ($95\%$C.L.) \\
\hline
$\Omega_{m}$                   & 0.300$\pm$0.017          & 0.302$\pm$0.010         & 0.3005$\pm$0.0092        & 0.291$\pm$0.015           & 0.2982$\pm$0.0099      & 0.2944$\pm$0.0093 \\
$H_0 $                         & 68.51$\pm$1.30           & 68.35$\pm$0.81          & 68.41$\pm$0.72           & 69.04$\pm$1.18            & 68.50$\pm$0.80          & 68.89$\pm$0.75 \\
\hline
\hline
$\chi^2_{\rm min}/2$           & 4900.765                 & 4901.399                & 4907.901                 & 4900.656                  & 4901.140                & 4908.286 \\
\hline
\end{tabular}
}
\caption{Constraints on cosmological parameters, using different combinations of CMB data sets, of linear \textit{pure} EFT (a) and designer $f(R)$ (b) models, on both $\Lambda$CDM (left) and $w$CDM (right) background.  \label{Tab:stat}}
\end{table*}

\clearpage



\begin{thebibliography}{99}

\bibitem{Bertschinger:2006aw} 
  E.~Bertschinger,
  Astrophys.\ J.\  {\bf 648}, 797 (2006),
  [astro-ph/0604485].
  
\bibitem{Linder:2007hg} 
  E.~V.~Linder and R.~N.~Cahn,
  Astropart.\ Phys.\  {\bf 28}, 481 (2007),
 [astro-ph/0701317].
  
\bibitem{Zhang:2007nk} 
  P.~Zhang, M.~Liguori, R.~Bean and S.~Dodelson,
  Phys.\ Rev.\ Lett.\  {\bf 99}, 141302 (2007),
  [arXiv:0704.1932 [astro-ph]].

\bibitem{Hu:2007pj} 
  W.~Hu and I.~Sawicki,
  Phys.\ Rev.\ D {\bf 76}, 104043 (2007),
  [arXiv:0708.1190 [astro-ph]].
  
\bibitem{Amendola:2007rr} 
  L.~Amendola, M.~Kunz and D.~Sapone,
  JCAP {\bf 0804}, 013 (2008),
  [arXiv:0704.2421 [astro-ph]].
  
\bibitem{Bertschinger:2008zb} 
  E.~Bertschinger and P.~Zukin,
  Phys.\ Rev.\ D {\bf 78}, 024015 (2008)
  [arXiv:0801.2431 [astro-ph]].

\bibitem{Daniel:2008et} 
  S.~F.~Daniel, R.~R.~Caldwell, A.~Cooray and A.~Melchiorri,
  Phys.\ Rev.\ D {\bf 77}, 103513 (2008),
  [arXiv:0802.1068 [astro-ph]].

\bibitem{Song:2008vm} 
  Y.~-S.~Song and K.~Koyama,
  JCAP {\bf 0901}, 048 (2009),
 [arXiv:0802.3897 [astro-ph]].

\bibitem{Skordis:2008vt} 
  C.~Skordis,
  Phys.\ Rev.\ D {\bf 79}, 123527 (2009),
  [arXiv:0806.1238 [astro-ph]].
  
\bibitem{Song:2008xd} 
  Y.~-S.~Song and O.~Dore,
  JCAP {\bf 0903}, 025 (2009),
  [arXiv:0812.0002 [astro-ph]].

\bibitem{Zhao:2009fn} 
  G.~-B.~Zhao {\it et al.}, 
  Phys.\ Rev.\ Lett.\  {\bf 103}, 241301 (2009),
 [arXiv:0905.1326 [astro-ph.CO]].

\bibitem{Song:2010rm} 
  Y.~-S.~Song, L.~Hollenstein, G.~Caldera-Cabral and K.~Koyama,
  JCAP {\bf 1004}, 018 (2010),
 [arXiv:1001.0969 [astro-ph.CO]].

\bibitem{Daniel:2010ky} 
  S.~F.~Daniel {\it et al.}, 
  Phys.\ Rev.\ D {\bf 81}, 123508 (2010),
  [arXiv:1002.1962 [astro-ph.CO]].

\bibitem{Pogosian:2010tj} 
  L.~Pogosian, A.~Silvestri, K.~Koyama and G.~-B.~Zhao,
  Phys.\ Rev.\ D {\bf 81}, 104023 (2010),
  [arXiv:1002.2382 [astro-ph.CO]].

\bibitem{Bean:2010zq} 
  R.~Bean and M.~Tangmatitham,
  Phys.\ Rev.\ D {\bf 81}, 083534 (2010),
  [arXiv:1002.4197 [astro-ph.CO]].

\bibitem{Zhao:2010dz} 
  G.~-B.~Zhao {\it et al.}, 
  Phys.\ Rev.\ D {\bf 81}, 103510 (2010),
  [arXiv:1003.0001 [astro-ph.CO]].
  
\bibitem{Dossett:2010gq} 
  J.~Dossett, M.~Ishak, J.~Moldenhauer, Y.~Gong, A.~Wang, Y.~Gong and A.~Wang,
  JCAP {\bf 1004}, 022 (2010)
  [arXiv:1004.3086 [astro-ph.CO]].
  
\bibitem{Thomas:2011pj} 
  S.~A.~Thomas, S.~A.~Appleby and J.~Weller,
  JCAP {\bf 1103}, 036 (2011),
  [arXiv:1101.0295 [astro-ph.CO]].

\bibitem{Baker:2011jy} 
  T.~Baker, P.~G.~Ferreira, C.~Skordis and J.~Zuntz,
  Phys.\ Rev.\ D {\bf 84}, 124018 (2011),
  [arXiv:1107.0491 [astro-ph.CO]].
  
\bibitem{Thomas:2011sf} 
  D.~B.~Thomas and C.~R.~Contaldi,
  JCAP {\bf 1112}, 013 (2011),
  [arXiv:1107.0727 [astro-ph.CO]].
  
\bibitem{Dossett:2011tn} 
  J.~N.~Dossett, M.~Ishak and J.~Moldenhauer,
  Phys.\ Rev.\ D {\bf 84}, 123001 (2011),
  [arXiv:1109.4583 [astro-ph.CO]].

\bibitem{Zhao:2011te} 
  G.~-B.~Zhao {\it et al.}, 
  Phys.\ Rev.\ D {\bf 85}, 123546 (2012),
  [arXiv:1109.1846 [astro-ph.CO]].

\bibitem{Hojjati:2011xd} 
  A.~Hojjati {\it et al.},	
  Phys.\ Rev.\ D {\bf 85}, 043508 (2012),
 [arXiv:1111.3960 [astro-ph.CO]].
  
\bibitem{Brax:2011aw} 
  P.~Brax, A.~-C.~Davis and B.~Li,
  Phys.\ Lett.\ B {\bf 715}, 38 (2012),
  [arXiv:1111.6613 [astro-ph.CO]].
  
\bibitem{Dossett:2012kd} 
  J.~Dossett and M.~Ishak,
  Phys.\ Rev.\ D {\bf 86}, 103008 (2012)
  [arXiv:1205.2422 [astro-ph.CO]].
  
\bibitem{Brax:2012gr}
  P.~Brax, A.~-C.~Davis, B.~Li and H.~A.~Winther,
  Phys.\ Rev.\ D {\bf 86}, 044015 (2012),
  [arXiv:1203.4812 [astro-ph.CO]].
  
\bibitem{Sawicki:2012re} 
  I.~Sawicki, I.~D.~Saltas, L.~Amendola and M.~Kunz,
  JCAP {\bf 1301}, 004 (2013),
  [arXiv:1208.4855 [astro-ph.CO]].
  
\bibitem{Baker:2012zs} 
  T.~Baker, P.~G.~Ferreira and C.~Skordis,
  Phys.\ Rev.\ D {\bf 87}, 024015 (2013),
  [arXiv:1209.2117 [astro-ph.CO]].
 
\bibitem{Amendola:2012ky} 
  L.~Amendola {\it et al.},  
  Phys.\ Rev.\ D {\bf 87}, 023501 (2013),
  [arXiv:1210.0439 [astro-ph.CO]].

\bibitem{Hojjati:2012ci} 
  A.~Hojjati,
  JCAP {\bf 1301}, 009 (2013),
  [arXiv:1210.3903 [astro-ph.CO]].
  
\bibitem{Silvestri:2013ne} 
  A.~Silvestri, L.~Pogosian and R.~V.~Buniy,
  Phys.\ Rev.\ D {\bf 87}, no. 10, 104015 (2013)
  [arXiv:1302.1193 [astro-ph.CO]].

\bibitem{Hu:2012td} 
  B.~Hu, M.~Liguori, N.~Bartolo and S.~Matarrese,
  Phys.\ Rev.\ D {\bf 88}, no. 2, 024012 (2013)
  [arXiv:1211.5032 [astro-ph.CO]].

\bibitem{Dossett:2013npa} 
  J.~Dossett and M.~Ishak,
  Phys.\ Rev.\ D {\bf 88}, 103008 (2013),
  [arXiv:1311.0726 [astro-ph.CO]].

\bibitem{Hu:2013aqa} 
  B.~Hu, M.~Liguori, N.~Bartolo and S.~Matarrese,
  Phys.\ Rev.\ D {\bf 88}, 123514 (2013),
  [arXiv:1307.5276 [astro-ph.CO]].

\bibitem{Motta:2013cwa} 
  M.~Motta, I.~Sawicki, I.~D.~Saltas, L.~Amendola and M.~Kunz,
  Phys.\ Rev.\ D {\bf 88}, 124035 (2013),
  [arXiv:1305.0008 [astro-ph.CO]].
  
\bibitem{Asaba:2013xql} 
  S.~Asaba {\it et al.},  
  JCAP {\bf 1308}, 029 (2013),
  [arXiv:1306.2546 [astro-ph.CO]].
  


	
\bibitem{Terukina:2013eqa} 
  A.~Terukina, L.~Lombriser, K.~Yamamoto, D.~Bacon, K.~Koyama and R.~C.~Nichol,
  arXiv:1312.5083 [astro-ph.CO].
	
\bibitem{Piazza:2013pua} 
  F.~Piazza, H.~Steigerwald and C.~Marinoni,
  arXiv:1312.6111 [astro-ph.CO].

\bibitem{Baker:2013hia} 
  T.~Baker, P.~G.~Ferreira and C.~Skordis,
  Phys.\ Rev.\ D {\bf 89}, 024026 (2014),
  [arXiv:1310.1086 [astro-ph.CO]].

\bibitem{Gergely:2014rna} 
  L\'as.~\'A.Gergely and S.~Tsujikawa,
  Phys.\ Rev.\ D {\bf 89}, 064059 (2014)
  [arXiv:1402.0553 [hep-th]].
  

\bibitem{Munshi:2014tua} 
  D.~Munshi, B.~Hu, A.~Renzi, A.~Heavens and P.~Coles,
  arXiv:1403.0852 [astro-ph.CO].
  


\bibitem{Steigerwald:2014ava} 
  H.~Steigerwald, J.~Bel and C.~Marinoni,
  arXiv:1403.0898 [astro-ph.CO].
	
\bibitem{Huang:2014fua} 
  Q.~-G.~Huang,
  arXiv:1403.0655 [astro-ph.CO].

\bibitem{Lombriser:2014dua} 
  L.~Lombriser,
  arXiv:1403.4268 [astro-ph.CO].

\bibitem{Tsujikawa:2014mba} 
  S.~Tsujikawa,
  arXiv:1404.2684 [gr-qc].


\bibitem{Bellini:2014fua} 
  E.~Bellini and I.~Sawicki,
  arXiv:1404.3713 [astro-ph.CO].
  
  \bibitem{planck}
http://sci.esa.int/planck\,.

\bibitem{sdss}
http://www.sdss.org\,.

 \bibitem{des} 
http://www.darkenergysurvey.org\,.

\bibitem{lsst} 
http://www.lsst.org\,.

\bibitem{euclid}
http://sci.esa.int/euclid\,.  

\bibitem{Drinkwater:2009sd} 
  M.~J.~Drinkwater, R.~J.~Jurek, C.~Blake, D.~Woods, K.~A.~Pimbblet, K.~Glazebrook, R.~Sharp and M.~B.~Pracy {\it et al.},
  Mon.\ Not.\ Roy.\ Astron.\ Soc.\  {\bf 401}, 1429 (2010)
  [arXiv:0911.4246 [astro-ph.CO]].

\bibitem{Parkinson:2012vd} 
  D.~Parkinson, S.~Riemer-Sorensen, C.~Blake, G.~B.~Poole, T.~M.~Davis, S.~Brough, M.~Colless and C.~Contreras {\it et al.},
  Phys.\ Rev.\ D {\bf 86}, 103518 (2012)
  [arXiv:1210.2130 [astro-ph.CO]].

\bibitem{Simpson:2012ra} 
  F.~Simpson, C.~Heymans, D.~Parkinson, C.~Blake, M.~Kilbinger, J.~Benjamin, T.~Erben and H.~Hildebrandt {\it et al.},
  arXiv:1212.3339 [astro-ph.CO].
   
\bibitem{Heymans:2013fya} 
  C.~Heymans, E.~Grocutt, A.~Heavens, M.~Kilbinger, T.~D.~Kitching, F.~Simpson, J.~Benjamin and T.~Erben {\it et al.},
  arXiv:1303.1808 [astro-ph.CO].
  
\bibitem{Fu:2014loa} 
  L.~Fu, M.~Kilbinger, T.~Erben, C.~Heymans, H.~Hildebrandt, H.~Hoekstra, T.~D.~Kitching and Y.~Mellier {\it et al.},
  arXiv:1404.5469 [astro-ph.CO].
 
  
\bibitem{Gubitosi:2012hu} 
  G.~Gubitosi, F.~Piazza and F.~Vernizzi,
  JCAP {\bf 1302}, 032 (2013)
  [arXiv:1210.0201 [hep-th]].
  
\bibitem{Bloomfield:2012ff} 
  J.~K.~Bloomfield, \'E. \'E.~Flanagan, M.~Park and S.~Watson,
  JCAP {\bf 1308}, 010 (2013)
  [arXiv:1211.7054 [astro-ph.CO]].
 
\bibitem{Battye:2012eu} 
  R.~A.~Battye and J.~A.~Pearson,
JCAP {\bf 1207}, 019 (2012), 
  [arXiv:1203.0398 [hep-th]]. 


\bibitem{Weinberg:2008hq} 
  S.~Weinberg,
Phys.\ Rev.\ D {\bf 77}, 123541  (2008), 
  [arXiv:0804.4291 [hep-th]]. 

\bibitem{Creminelli:2008wc} 
  P.~Creminelli, G.~D'Amico, J.~Norena and F.~Vernizzi,
  JCAP {\bf 0902}, 018 (2009),
   [arXiv:0811.0827 [astro-ph]].
  
\bibitem{Park:2010cw} 
  M.~Park, K.~M.~Zurek and S.~Watson,
  Phys.\ Rev.\ D {\bf 81}, 124008 (2010),
   [arXiv:1003.1722 [hep-th]].
 
\bibitem{Cheung:2007st} 
  C.~Cheung {\it et al.}, 
  JHEP {\bf 0803}, 014 (2008),
  [arXiv:0709.0293 [hep-th]].
  
\bibitem{Jimenez:2011nn} 
  R.~Jimenez, P.~Talavera and L.~Verde,
  Int.\ J.\ Mod.\ Phys.\ A {\bf 27}, 1250174 (2012),
  [arXiv:1107.2542 [astro-ph.CO]].
  
\bibitem{Carrasco:2012cv} 
  J.~J.~M.~Carrasco, M.~P.~Hertzberg and L.~Senatore,
JHEP {\bf 1209}, 082  (2012), 
 [arXiv:1206.2926 [astro-ph.CO]].  
 
\bibitem{Hertzberg:2012qn} 
  M.~P.~Hertzberg,
  Phys.\ Rev.\ D {\bf 89}, 043521 (2014),
  [arXiv:1208.0839 [astro-ph.CO]].

\bibitem{Gleyzes:2013ooa} 
  J.~Gleyzes, D.~Langlois, F.~Piazza and F.~Vernizzi,
  JCAP {\bf 1308}, 025 (2013),
  [arXiv:1304.4840 [hep-th]].
	
\bibitem{Bloomfield:2013efa} 
  J.~Bloomfield,
  JCAP {\bf 12}, 044 (2013),
  [arXiv:1304.6712 [astro-ph.CO]].
  
\bibitem{Piazza:2013coa} 
  F.~Piazza and F.~Vernizzi,
  Class.\ Quant.\ Grav.\  {\bf 30}, 214007 (2013)
  [arXiv:1307.4350].
  
\bibitem{Hu:2013twa} 
  B.~Hu, M.~Raveri, N.~Frusciante and A.~Silvestri,
  arXiv:1312.5742 [astro-ph.CO].
  
   \bibitem{CAMB}
http://camb.info \,.

\bibitem{Lewis:1999bs} 
  A.~Lewis, A.~Challinor and A.~Lasenby,
  Astrophys.\ J.\  {\bf 538}, 473 (2000),
  [astro-ph/9911177].
  
\bibitem{Lewis:2002ah}
  A.~Lewis and S.~Bridle,
  Phys.\ Rev.\ D {\bf 66}, 103511 (2002)
  [astro-ph/0205436].
  
\bibitem{HRFS-work-in-progress} 
 M.~Raveri, B.~Hu, N.~Frusciante, A.~Silvestri,
  ``EFTCAMB/EFTCosmoMC: Numerical Notes v1.0' 
\emph{in preparation}

\bibitem{Horndeski:1974wa} 
  G.~W.~Horndeski,
  Int.\ J.\ Theor.\ Phys.\  {\bf 10}, 363 (1974).

\bibitem{Copeland:2006wr} 
  E.~J.~Copeland, M.~Sami and S.~Tsujikawa,
  Int.\ J.\ Mod.\ Phys.\ D {\bf 15}, 1753 (2006)
  [hep-th/0603057].

\bibitem{ArmendarizPicon:1999rj} 
  C.~Armendariz-Picon, T.~Damour and V.~F.~Mukhanov,
  Phys.\ Lett.\ B {\bf 458}, 209 (1999)
  [hep-th/9904075].

\bibitem{Deffayet:2009wt} 
  C.~Deffayet, G.~Esposito-Farese and A.~Vikman,
  Phys.\ Rev.\ D {\bf 79}, 084003 (2009)
  [arXiv:0901.1314 [hep-th]].

\bibitem{Nicolis:2008in} 
  A.~Nicolis, R.~Rattazzi and E.~Trincherini,
  Phys.\ Rev.\ D {\bf 79}, 064036 (2009),
  [arXiv:0811.2197 [hep-th]].

\bibitem{Bloomfield:2013cyf} 
  J.~Bloomfield and J.~Pearson,
  JCAP {\bf 1403}, 017 (2014),
  [arXiv:1310.6033 [astro-ph.CO]].

\bibitem{Battye:2013ida} 
  R.~A.~Battye and J.~A.~Pearson,
  JCAP {\bf 1403}, 051 (2014),
  [arXiv:1311.6737 [astro-ph.CO]].
  
\bibitem{Frusciante:2013zop} 
  N.~Frusciante, M.~Raveri and A.~Silvestri,
  JCAP {\bf 1402}, 026 (2014),
  [arXiv:1310.6026 [astro-ph.CO]].
  
\bibitem{Chevallier:2000qy} 
  M.~Chevallier and D.~Polarski,
  Int.\ J.\ Mod.\ Phys.\ D {\bf 10}, 213 (2001),
 [gr-qc/0009008].
	
\bibitem{Linder:2002et} 
  E.~V.~Linder,
  Phys.\ Rev.\ Lett.\  {\bf 90}, 091301 (2003),
  [astro-ph/0208512].
  
\bibitem{Song:2006ej} 
  Y.~-S.~Song, W.~Hu and I.~Sawicki,
  Phys.\ Rev.\ D {\bf 75}, 044004 (2007),
  [astro-ph/0610532].
	
\bibitem{Pogosian:2007sw} 
  L.~Pogosian and A.~Silvestri,
  Phys.\ Rev.\ D {\bf 77}, 023503 (2008),
  [Erratum-ibid.\ D {\bf 81}, 049901 (2010)],
  [arXiv:0709.0296 [astro-ph]].
  
\bibitem{Gilks:1999}
 	W. R. Gilks,
 	 ``Markov Chain Monte Carlo In Practice'', 
	Chapman and Hall/CRC, (1999), 0412055511.
  
\bibitem{Ade:2013kta}
  P.~A.~R.~Ade {\it et al.}  [Planck Collaboration],
  arXiv:1303.5075 [astro-ph.CO].

\bibitem{Ade:2013zuv}
  P.~A.~R.~Ade {\it et al.}  [Planck Collaboration],
  arXiv:1303.5076 [astro-ph.CO].

\bibitem{Ade:2013tyw} 
  P.~A.~R.~Ade {\it et al.}  [Planck Collaboration],
  arXiv:1303.5077 [astro-ph.CO].
  
  
\bibitem{Hinshaw:2012aka}
  G.~Hinshaw {\it et al.}  [WMAP Collaboration],
  Astrophys.\ J.\ Suppl.\  {\bf 208}, 19 (2013),
  [arXiv:1212.5226 [astro-ph.CO]].
  
\bibitem{Beutler:2011hx} 
  F.~Beutler, C.~Blake, M.~Colless, D.~H.~Jones, L.~Staveley-Smith, L.~Campbell, Q.~Parker and W.~Saunders {\it et al.},
  Mon.\ Not.\ Roy.\ Astron.\ Soc.\  {\bf 416}, 3017 (2011),
  [arXiv:1106.3366 [astro-ph.CO]].
  
\bibitem{Percival:2009xn} 
  W.~J.~Percival {\it et al.}  [SDSS Collaboration],
  Mon.\ Not.\ Roy.\ Astron.\ Soc.\  {\bf 401}, 2148 (2010),
  [arXiv:0907.1660 [astro-ph.CO]].
  
\bibitem{Padmanabhan:2012hf} 
  N.~Padmanabhan, X.~Xu, D.~J.~Eisenstein, R.~Scalzo, A.~J.~Cuesta, K.~T.~Mehta and E.~Kazin,
  Mon.\ Not.\ Roy.\ Astron.\ Soc.\  {\bf 427}, no. 3, 2132 (2012),
  [arXiv:1202.0090 [astro-ph.CO]].
  
\bibitem{Anderson:2012sa} 
  L.~Anderson, E.~Aubourg, S.~Bailey, D.~Bizyaev, M.~Blanton, A.~S.~Bolton, J.~Brinkmann and J.~R.~Brownstein {\it et al.},
  Mon.\ Not.\ Roy.\ Astron.\ Soc.\  {\bf 427}, no. 4, 3435 (2013),
  [arXiv:1203.6594 [astro-ph.CO]].
  
\bibitem{Zlatev:1998tr} 
  I.~Zlatev, L.~-M.~Wang and P.~J.~Steinhardt,
  Phys.\ Rev.\ Lett.\  {\bf 82}, 896 (1999)
  [astro-ph/9807002].

\bibitem{Bean:2006up} 
  R.~Bean, D.~Bernat, L.~Pogosian, A.~Silvestri and M.~Trodden,
  Phys.\ Rev.\ D {\bf 75}, 064020 (2007),
  [astro-ph/0611321].
  
\bibitem{DeFelice:2010aj} 
  A.~De Felice and S.~Tsujikawa,
  Living Rev.\ Rel.\  {\bf 13}, 3 (2010),
  [arXiv:1002.4928 [gr-qc]].
  
\bibitem{Starobinsky:2007hu} 
  A.~A.~Starobinsky,
  JETP Lett.\  {\bf 86}, 157 (2007),
  [arXiv:0706.2041 [astro-ph]].
  
\bibitem{Amendola:2006we} 
  L.~Amendola, R.~Gannouji, D.~Polarski and S.~Tsujikawa,
  Phys.\ Rev.\ D {\bf 75}, 083504 (2007)
  [gr-qc/0612180].
  
\bibitem{Hu:2007nk} 
  W.~Hu and I.~Sawicki,
  Phys.\ Rev.\ D {\bf 76}, 064004 (2007)
  [arXiv:0705.1158 [astro-ph]].

\bibitem{Zhao:2008bn} 
  G.~-B.~Zhao, L.~Pogosian, A.~Silvestri and J.~Zylberberg,
  Phys.\ Rev.\ D {\bf 79}, 083513 (2009)
  [arXiv:0809.3791 [astro-ph]].
  
\bibitem{Hojjati:2011ix} 
  A.~Hojjati, L.~Pogosian and G.~-B.~Zhao,
  JCAP {\bf 1108}, 005 (2011)
  [arXiv:1106.4543 [astro-ph.CO]].
  
\bibitem{Hojjati:2012rf} 
  A.~Hojjati, L.~Pogosian, A.~Silvestri and S.~Talbot,
  Phys.\ Rev.\ D {\bf 86}, 123503 (2012),
  [arXiv:1210.6880 [astro-ph.CO]].

\bibitem{Dossett:2014oia} 
  J.~Dossett, B.~Hu and D.~Parkinson,
  JCAP {\bf 1403}, 046 (2014),
  [arXiv:1401.3980 [astro-ph.CO]].


  


\end{thebibliography}
\end{document}